\documentclass{emulateapj}
\usepackage{epstopdf}

\usepackage{natbib}

\newcommand{\ms}{M$_{\odot}$}
\newcommand{\ls}{L$_{\odot}$}
\newcommand{\kms}{$\,\rm km\,s^{-1}$}
\newcommand{\kkms}{$\,\mathrm{K\,km\,s^{-1}}$}

\newcommand{\hh}{H$_2$}

\newcommand{\ci}{[CI]}
\newcommand{\eqq}{\!=\!}  
\newcommand{\too}{\!\rightarrow\!} 
\newcommand{\jone}{{$J\eqq 1\too0$}}
\newcommand{\jtwo}{{$J\eqq2\too1$}}
\newcommand{\jthree}{{$J\eqq3\too2$}}
\newcommand{\jfour}{{$J\eqq4\too3$}}
\newcommand{\jfive}{{$J\eqq5\too4$}}
\newcommand{\jsix}{{$J\eqq6\too5$}}
\newcommand{\jseven}{{$J\eqq7\too6$}}
\newcommand{\jeight}{{$J\eqq8\too7$}}
\newcommand{\jnine}{{$J\eqq9\too8$}}
\newcommand{\waterline}{$\rm 2_{0,2}\too1_{1,1}$}
\newcommand{\waterlinenon}{$\rm 2_{1,1}\too 2_{0,2}$}
\newcommand{\groundortho}{$\rm 1_{1,0}\too 1_{0,1}$}

\newcommand{\as}{$^{\prime\prime}$}
\newcommand{\mm}{$\rm \mu m$}

\newcommand{\water}{$\rm{H_2O}$}

\newcommand{\ttt}{$\tau_{\rm 225 GHz}$}
\newcommand{\jykms}{$\rm Jy\,km\,s^{-1}$}
\shorttitle{Warm Molecular Gas in the Cloverleaf}
\shortauthors{Bradford et al.}
\begin{document}
\bibliographystyle{apj}
\title{The Warm Molecular Gas Around the Cloverleaf Quasar}

\slugcomment{Astrophysical Journal, in press}

\author{C.M. Bradford\altaffilmark{1,2}, J.E. Aguirre\altaffilmark{3,4}, R. Aikin\altaffilmark{3,2}, J.J. Bock\altaffilmark{1,2}, L. Earle\altaffilmark{3}, J. Glenn\altaffilmark{3}, H. Inami\altaffilmark{5}, P.R. Maloney\altaffilmark{3}, H. Matsuhara\altaffilmark{5}, B.J. Naylor\altaffilmark{2,1}, H.T. Nguyen\altaffilmark{1}, J. Zmuidzinas\altaffilmark{2,1}}

\altaffiltext{1}{Jet Propulsion Laboratory, Pasadena, CA, 91109}
\altaffiltext{2}{California Institute of Technology, Pasadena, CA, 91125}
\altaffiltext{3}{University of Colorado, Boulder, CO, 80303}
\altaffiltext{4}{University of Pennsylvania, Philadelphia, PA, 19104}
\altaffiltext{5}{Institute for Space and Astronautical Science, Japan Aerospace and Exploration Agency, Sagamihara, Japan}

\begin{abstract}
We present the first broadband $\lambda = 1\rm mm$ spectrum toward the $z=2.56$ Cloverleaf Quasar, obtained with Z-Spec, a 1-mm grating spectrograph on the 10.4-meter Caltech Submillimeter Observatory. The 190--305~GHz observation band corresponds to rest-frame 272 to 444~\mm, and we measure the dust continuum as well as all four transitions of carbon monoxide (CO) lying in this range.  The power-law dust emission, $F_\nu = 14\rm mJy \left({\nu}/{240 GHz}\right)^{3.9}$ is consistent with the published continuum measurements.  The CO \jsix, \jeight, and  \jnine\ measurements are the first, and now provide the highest-J CO information in this source.  Our measured CO intensities are very close to the previously-published interferometric measurements of \jseven, and we use all available transitions and our $^{13}$CO upper limits to constrain the physical conditions in the Cloverleaf molecular gas disk.  We find a large mass (2--50$\times10^{9}\,\rm M_{\odot}$) of highly-excited gas with thermal pressure $\rm nT > 10^6\,\rm K cm^{-3}$.  The ratio of the total CO cooling to the far-IR dust emission exceeds that in the local dusty galaxies, and we investigate the potential heating sources for this bulk of warm molecular gas.  We conclude that both UV photons and X-rays likely contribute, and discuss implications for a top-heavy stellar initial mass function arising in the X-ray-irradiated starburst.   Finally we present tentative identifications of other species in the spectrum, including a possible detection of the \water\ \waterline\ transition at $\rm \lambda_{rest}$ = 303~\mm.  

\end{abstract}


\section{Introduction}
Since the first measurements of CO in the powerful IRAS galaxy FSC10214 at z=2.3~\citep{bvb92} nearly two decades ago,  the field of molecular line measurements in early-universe galaxies has grown steadily (see review by \citet{svb05}).  Several tens of sources are now known, drawn from a wide variety of parent populations including IRAS galaxies, optically-selected quasars, submillimeter galaxies, and now Spitzer-selected galaxies (e.g. \cite{frayer08}).  As with the submillimeter and millimeter-wave continuum measurements, observations of high-z molecular lines benefit from a negative K-correction: the spectral lines generally carry more power at higher frequency.   In particular for carbon monoxide, the run of the CO line luminosity with J typically increases up to the mid-J transitions (J$\sim$5--10) for actively starforming galaxies.  This CO rotational spectrum when measured in energy units thus peaks in the 200--500~\mm\ regime, 
making millimeter-wave spectroscopy a powerful tool for studying molecular gas in the early universe.   Measurements of the CO spectrum across its peak constrains the temperature, density and total mass of molecular gas.  While observations of mid-J CO transitions in the local universe are hampered by poor transmission in narrow atmospheric windows,  a millimeter-wave systems can have approximately uniform sensitivity across the peak of the CO SED for $z>$1.   

All high-$z$ millimeter-wave observations to date have employed heterodyne receivers, originally with single-dish telescopes but soon after with large interferometer arrays.  SIS receiver sensitivities, $T_{rec}\sim \rm 30~K$, are close to the photon background limit at mountaintop sites, and are an excellent match to the large-collecting area arrays.
However, their instantaneous fractional bandwidths are at most $\sim10\%$, even with the new wideband backends (e.g. 8~GHz for ALMA).   They thus require a separate tuning for each spectral line, and searching for unknown lines and/or unknown redshifts is time-consuming.   A more comprehensive approach to finding redshifts and probing molecular gas contents of distant galaxies can be obtained with complete spectral coverage, to the extent possible given the telluric windows.  CO is but one of the important coolants, and it is generally accepted that other species, notably water, become important as the molecular gas becomes more excited.   Z-Spec is the first spectrograph designed specifically for this type of measurement at wavelengths longward of the mid-IR; a first-order grating with simultaneous coverage over the entire 1 mm atmospheric window.  

After a brief introduction to the Cloverleaf quasar, Section~\ref{sec:obs} provides a description of Z-Spec and our observations at the CSO.   The line and continuum parameters extracted from our spectrum are presented in Section~\ref{sec:results}, and the findings are discussed in Section~\ref{sec:discussion}.   

\begin{figure*}[t!]\centering
\includegraphics*[width=16cm,bb=53 5 670 440]{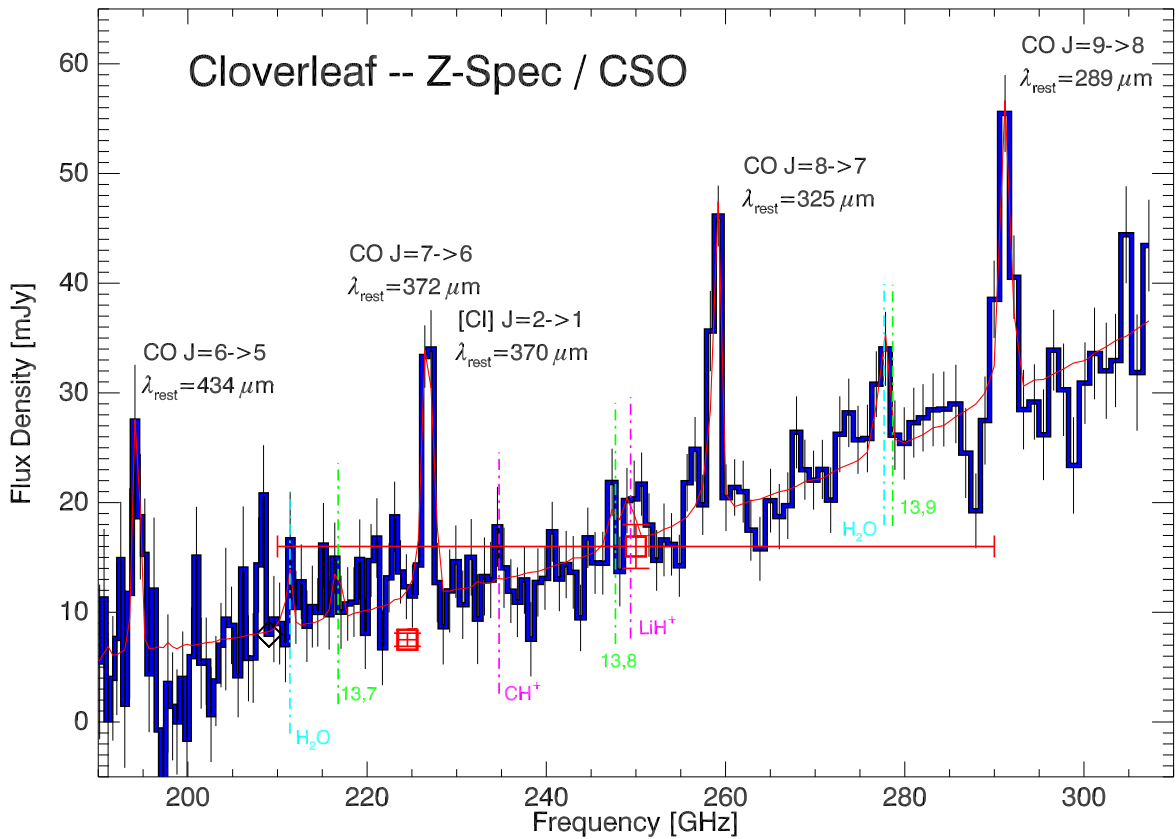}
\caption{ Z-Spec Spectrum toward the z=2.56 Cloverleaf Quasar.   Only one channel of the total 160 has been removed due to high noise, at 209.0~GHz---a value near the baseline is plotted in its place, noted with a diamond.  A fit to the continuum and six spectral lines is overlaid with a light red curve.  Red squares with error bars and spectral widths denote continuum measurements from MAMBO and Plateau de Bure (Section~\ref{sec:continuum}).  Thin dot-dash lines mark transitions frequencies for which upper limits are extracted .  The \jseven, \jeight, and \jnine\ $^{13}$CO frequencies are marked in green with 13,7, 13,8, and 13,9.  Frequencies of the two water transitions with $E_{\rm upper}<140~\rm K$ in the band are marked in cyan.  The LiH and CH$^+$ frequencies and a tentative absorption feature are are marked with magenta.  Upper limits are provided for all transitions except $^13$CO~\jnine\ in Table~\ref{tab:lines}.  See also section~\ref{sec:limits}. \label{fig:spec}}
\end{figure*}

\subsection{The Cloverleaf}  We have observed the lensed $\rm z=2.56$ broad-absorption-line quasar system H1413+1143, also known as the Cloverleaf.   The source was originally detected in an optical spectroscopic survey \citep{Hazard_84}, but a warm dust component also emits some $\sim 7\times 10^{13}$\ls\ (intrinsic) in the rest-frame mid-IR \citep
{Lutz_07}, ascribed to reradiated energy from the accretion zone.   The SED also suggests a second distinct dust emission component dominating the rest-frame far-IR and submillimeter, but with a total bolometric luminosity $\sim$5--10\% that of the AGN dust component \citep{Barvainis_92,Barvainis_95,Weiss_05}.  The measurement of PAH features with intensities consistent with a starburst in this far-IR / submm component leads to the picture that the Cloverleaf is a composite object:  an usually powerful partially-embedded AGN accompanied by a weaker but still tremendous starbust which alone would be similar to a submillimeter galaxy~\citep{Lutz_07}.   These large luminosities, combined with a substantial lensing magnification factor ($\sim11$) makes the Cloverleaf an excellent laboratory for studying the AGN / host interaction in what is likely the era of peak activity in galaxies.  In particular, the molecular gas reservoir is the raw material for both the star formation and ultimately the nuclear accretion, and the impact of UV and X-ray photons from the stars or AGN on the molecular gas is a key aspect of the starburst / AGN interaction.

Some of the earliest high-$z$ molecular gas measurements were detections of the CO \jthree\ transition redshifted to the 3~mm band \citep{Barvainis_94, Barvainis_97}.  These measurements confirmed the presence of a large gas reservoir ($\sim 10^{10}$~\ms) inferred from the dust SED.  Subsequent observations of the \jthree\ and \jseven\ transitions with the BIMA, OVRO and IRAM interferometers have steadily improved \citep{Wilner_95, Yun_97}, culminating in an IRAM \jseven\ map which fully resolves the four Cloverleaf components (\citet{Alloin_97} [hereafter A97], \citet{Kneib_98a}).   \citet{Venturini_03} used the \jseven\ image to model the CO \jseven\ source and lens and find (after correcting to the modern cosmology) that the intrinsic source is a disk with radius of 650~pc, inclined by 30$^\circ$.  The lens is formed with two galaxies of comparable mass at 0.25\as\ and 0.71\as\ from the line of sight.  Our Z-Spec measurement complements this spatial information by providing a survey of several CO transitions spanning the peak of the spectrum.  The uniform calibration allows us to anchor the total molecular gas conditions and energetics.  The result provides constraints on the gas heating sources: stars or the active nucleus, and on the impact of the active nucleus on the putative host starburst.

\section{Z-Spec Instrument and Observations}\label{sec:obs}

Z-Spec is a single-beam grating spectrometer which disperses the 190--308~GHz band across an linear array of 160 bolometers.   The grating approach is novel: a curved grating operates in a parallel plate waveguide which is fed by a single-mode corrugated feedhorn.   The instrument coupling to the CSO telescope is thus well-approximated with a Gaussian-beam approach, and the measured efficiencies and beam sizes are consistent with this.   Details of the grating design and testing can be found in \citet{Naylor_03, Bradford_04,Earle_06, Inami_08}.   The detector spacing increases from $\sim1$ part in 400 at the low-frequency end ($\Delta\nu = 500$~MHz), to $\sim1$ part in 250 at the high-frequency end of the band ($\Delta\nu =1200$~MHz), while the spectrometer resolving power runs from $\sim1$ part in 300 at low frequencies to $\sim1$ part in 250 at the high frequency end of the band.   Thus the system is marginally under-sampled, especially at the high-frequencies.   

Spectral profiles for all 160 channels have been measured with a long-path Fourier-transform spectrometer, with channel center frequencies adjusted slightly per observations of multiple transitions in the spectral standard IRC+10216.  Refinement of the Z-Spec frequency scale is ongoing as we incorporate line measurements from an increasing library of astronomical sources; for the data presented here, we are confident that the channel frequencies are known to better than 200~MHz across the band, or approximately one fifth of a channel width.  We expect to improve this further in the near future.

The entire structure is cooled with an adiabatic demagnetization refrigerator and operates at temperatures between 60 and 85~mK to facilitate photon-background-limited detection.    The bolometers developed at the JPL Microdevices Laboratory are individually-mounted silicon-nitride micro-mesh absorbers with quarter-wave backshorts, read-out with neutron-transmutation-doped Germanium thermistors.  With an operational optical loading is $\sim1-3 \times 10^{-13}\rm W$, and phonon NEPs of $4\times10^{-18} \rm W Hz^{-1/2}$, these detectors are the most-sensitive, lowest-background bolometers fielded to date for astrophysics.

Z-Spec observes in a traditional chop-and-nod mode, with the secondary chopping between 1 and 2 Hz, and a nod period of 20 seconds.   Because it is not possible to modulate the spectral response of the instrument relative to the bolometer array, and the spectral resolution elements are not oversampled, it is critical to both insure excellent array yield, and carefully calibrate each detector's response.   The situation is complicated by the fact that the both the bolometer loading and the bath temperature are changing throughout a typical observing night, while primary calibration sources are relatively scarce.  To address this, we have built a library of planetary observations in varying conditions, and we fit the dependence of each bolometer's response on its operating voltage, a proxy for the combination of bath temperature and optical loading.  This relationship provides a calibration correction which is used to bridge intervals between astronomical calibration observations.   Based on the self-consistency obtained with this scheme on planets and quasar calibrators, we estimate that the channel-to-channel calibration uncertainties are less than 10\%, except at the lowest frequencies which are degraded by the wing of the 186~GHz atmospheric water line.  

The Cloverleaf was observed on 2008 April 8 and April 15, with a total observing time (including chopping) of 7.9 hours, equally split between the two nights.  On April 8, the weather was excellent, with \ttt\ between 0.04 and 0.06, while April 16 had higher opacity: \ttt $\sim$ 0.15--0.2.   Each channel's signal is summed individually on a nod-by-nod basis, with each nod individually calibrated and weighted by the inverse square of the detector noise (in Janskys) measured around the chop frequency.   No modifications are made to the overall spectral shape at any point.   Z-Spec's sensitivity is weather dependent:  on April 8, the sensitivity per spectral channel was 300--500~$\rm mJy\,\sqrt{sec}$, but degraded to 500--700~$\rm mJy\,\sqrt{sec}$ on the 15th, with greater degradation at the band edges, where the opacity increase is larger.  The measured sensitivities are consistent with a simple model in which photon noise from the sky and telescope is the dominant contribution to the system noise.   The final RMS uncertainty in each channel is shown with the error bars in Figure~1.  It ranges from 2.5 to 4 mJy for most channels, with a median of 3.3~mJy. 


\section{Results, Continuum and Line Flux Extractions}\label{sec:results}

The Z-Spec spectrum toward the Cloverleaf system is shown in Figure~\ref{fig:spec}.   To extract line fluxes, we perform a simultaneous fit to a single component power-law continuum and the multiple lines.  Each channel's measured spectral response profile is used in the fitting since the spectrometer is not critically sampled, and the grating has spectral sidelobes at the $\sim$1\% level.  
All line widths are fixed, since Z-Spec is not sensitive to widths below $\sim$1000~\kms.  The PdB measurements of the CO \jthree\ shows a width of 420~\kms (W03), while the \citet{Kneib_98a} spectrum of \jseven\ has a Gaussian fit to some of the profile giving FWHM$\sim$450~\kms, though the profile is non-Gaussian and the true RMS width of the profile is larger than this.  It is conceivable that the line width increases with $J$---this would be expected if warmer material lies preferentially closer to the point-like nucleus.  We adopt 500~\kms\ for the Z-Spec fits.  This choice is not crucial; we have found that the fits for the integrated line fluxes are not strongly sensitive to the adopted line width for values below $\sim$800~\kms.  A total of six spectral lines are fit, the results are presented in Table~\ref{tab:lines}.  

The \ci\ \jtwo\ ($\nu=227.5~\rm GHz$) and the CO~\jseven\ ($\nu=226.7~\rm GHz$) transitions are separated by only 1000~\kms, or about one Z-Spec channel, so blending is a problem.  To address this, we adopt an iterative approach\ to make appropriate use of prior information:   we first fit the continuum plus the three unblended CO lines (\jsix, \jeight, and \jnine) to determine their fluxes and a CO redshift.   We find $z=2.5585 \pm 0.0015$, a value in good agreement with previous measurements (our error derives from the accuracy of our frequency calibration, and can be improved in the future).   
With the redshift determined, we then fit fluxes to the CO~\jseven\ and \ci\ \jtwo\ lines, finding $45.3\pm6.4$ and $8.5\pm7.4$~\jykms, respectively (see Table~\ref{tab:lines}).  Our current frequency scale uncertainty corresponds to interchanging $\sim$4~\jykms\ of flux between the two transitions about these fit values, with the sum of the two conserved, a value which is less than but comparable to the statistical errors.
Our results are consistent with the various \jseven\ integrated intensity measurements from the Plateau de Bure.  
B97 and A97 quote $43.7\pm2.2$ and $43.3\pm2.4$~\jykms, respectively, using a fit to a Gaussian profile with $\Delta V_{\rm FWHM} =376$\kms\ (adopted based on the lower-$J$ measurements).  However, A97 also report 50.1$\pm2.8$ \jykms\ when the linewidth is fit as a free parameter (finding $\Delta V_{\rm FWHM} = 480$ \kms).   Given the linewidth discussion above, we adopt the larger of these as our input to the analysis, but adopt a 15\% systematic uncertainty to accommodate the range of measurements.   As the CO temperature spectrum plotted in  Figure~\ref{fig:sed} shows, this is in good agreement with the \jsix\ and \jeight\ measurements, giving confidence to the overall calibration and noise estimates.   Our \ci\ flux is also formally consistent with the W05 measurement ($5.2\pm0.3$~\jykms). 

An emission feature at 277.6~GHz is also fit, tentatively identified as water and discussed in Section~\ref{sec:water}.  Finally, we have obtained upper limits to six other transitions, discussed briefly in Section~\ref{sec:limits}.   The potential emission feature at 208~GHz is unidentified at present.

\begin{deluxetable*}{cccccccc}
\tabletypesize{\scriptsize}
\tablecaption{Far-IR and Submillimeter Transitions Observed in the Cloverleaf }
\tablewidth{0pt}
\tablehead{
\colhead{transition}&\colhead{$\nu_{\rm rest} {\rm[GHz]}$}&\colhead{E$_{\rm upper}$ [K]}&\colhead{$\nu_{\rm obs}$ [GHz]}&\colhead{flux [Jy km/s]}&\colhead{err [Jy km/s]\tablenotemark{a}}&\colhead{L [10$^8$ \ls]\tablenotemark{b}}&\colhead{ref\tablenotemark{c}}
}
\startdata
CO \jthree\ & 345.79 & 33.2 & 97.2 & 13.2 & 1.7 & 0.55 & W03\\
CO \jthree\ & 345.79 & 33.2 & 97.2 & 9.9 & 0.6 & 0.41 & B97\\
{[CI]} \jone & 492.16& 23.6 &138.3 &3.6&0.4 & 0.21 &W05\\
CO \jfour & 461.04 & 55.3 & 129.58 & 21.1 & 0.8 & 1.17 & B97\\
CO \jfive & 576.27 & 83.0& 161.96 & 24.0 & 1.7 & 1.66 & B97\\
CO \jsix & 691.47 & 116 &194.3 & 37.0& 8.1 & 3.1 & this work \\
CO \jseven & 806.65& 155 & 226.7& 50.1 &2.2 & 4.9& B97, A97 \\
CO \jseven & 806.65 & 155 & 226.7 & 45.3 & 6.3 & 4.4 & this work\\ 
{[CI]} \jtwo & 809.34 & 37 & 226.6 & 5.2 & 0.3 & 0.50 & W05 \\
{[CI]} \jtwo & 809.34 & 37 & 226.6 & 8.5 & 7.4 & 0.83 & this work\\
CO \jeight & 921.80& 199 & 259.0 & 51.4 & 4.7 & 5.7 & this work\\
CO \jnine & 1036.91 & 249& 291.4& 41.8 & 5.8 & 5.2 & this work \\
far-IR dust & \multicolumn{5}{c}{115~K, peaking at 100~\mm\ rest frame} & 5.4$\times 10^4$ & W03, L07\\
\\
\multicolumn{8}{c}{\underline{Tentative Identifications and 2.5 $\sigma$ Upper Limits}}\\
\water~\waterline & 987.93 & 101 & 277.6 & 20.3 & 6.1 & 2.4 \\
\water~\waterlinenon & 752.03	& 137 & 211.4 & $<$14.2 & 5.7 & $<$1.3\\ 
Absorption feature &  & & 288.3 & -17.4 & 5.8 \\
LiH \jtwo & 887.29 & 21.3 & 249.3 &  $<$12.3 & 4.9 & $<$1.9 \\
CH$^+$ \jone & 835.07 & 43.1 & 234.7 & $<$14.0 & 5.6 & $<$1.6  \\
$^{13}$CO \jseven & 771.19 & 148 & 216.81 & $<$11.7 & 4.7 & $<$1.1 \\
$^{13}$CO \jeight & 881.27 & 190 & 247.76 & $<$14.8 & 5.9 & $<$1.6 \\
\enddata
\tablenotetext{a}{Statistical uncertainties only, additional systematic uncertainties are included in the analysis.}
\tablenotetext{b}{Luminosities are corrected for lensing, using a magnification factor of 11.}  
\tablenotetext{c}{References --- W03: \citet{Weiss_03}, W05: \citet{Weiss_05}, B97: \citet{Barvainis_97}, A97: \citet{Alloin_97}, 
}
\label{tab:lines}
\end{deluxetable*}

\section{Discussion}\label{sec:discussion}

\subsection{Dust Continuum}\label{sec:continuum}
The best fit continuum across the Z-Spec band is $F_{\nu} =  14.1\pm 0.4\, \rm mJy \left({\nu}/{240 GHz}\right)^{3.91\pm0.17}$.   This is consistent with previous measurements, for which the best compilation is in W03 (their Figure~3); we show the latest  
measurements in our band in Figure~\ref{fig:spec}.   The MAMBO bolometer measurement lies very close to the Z-Spec spectrum; the PdB continuum measurement at 224~GHz is slightly lower, but we note that this data point lies below the best fit of W03 including all the higher-frequency data.  Indeed, the various measurements and limits from 97 GHz to 400 GHz do not all lie on a single greybody curve, but our best-fit exponent of  3.91 from 190 to 300 GHz is consistent with the W03 best fit: dust at 50~K with $\beta=2$ and with unit emissivity between 50--300~\mm. 

\begin{figure}\centering
\includegraphics*[width=8 cm,bb=50 40 430 440]{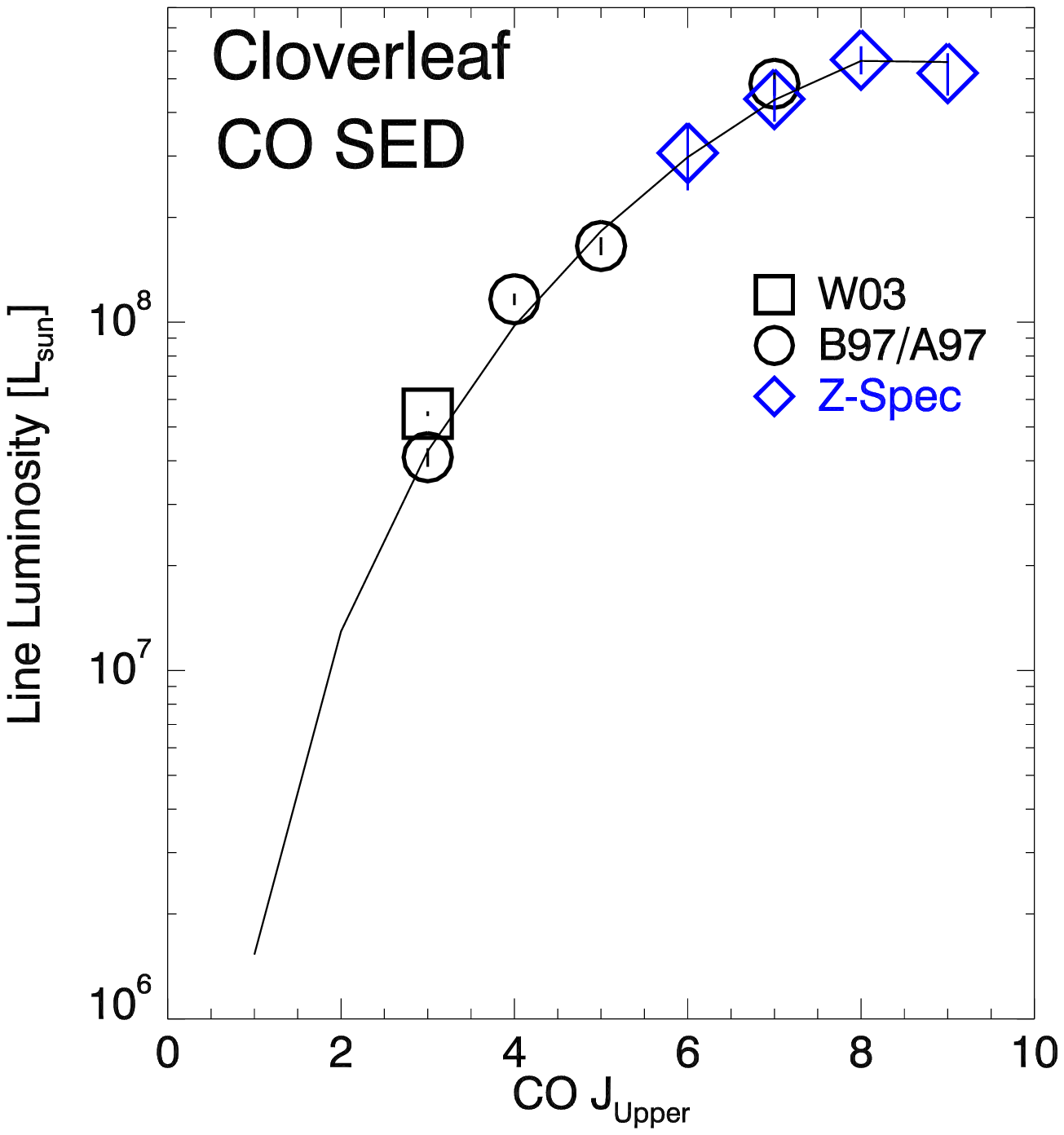}\\
\includegraphics*[width=8 cm,bb=50 5 430 264]{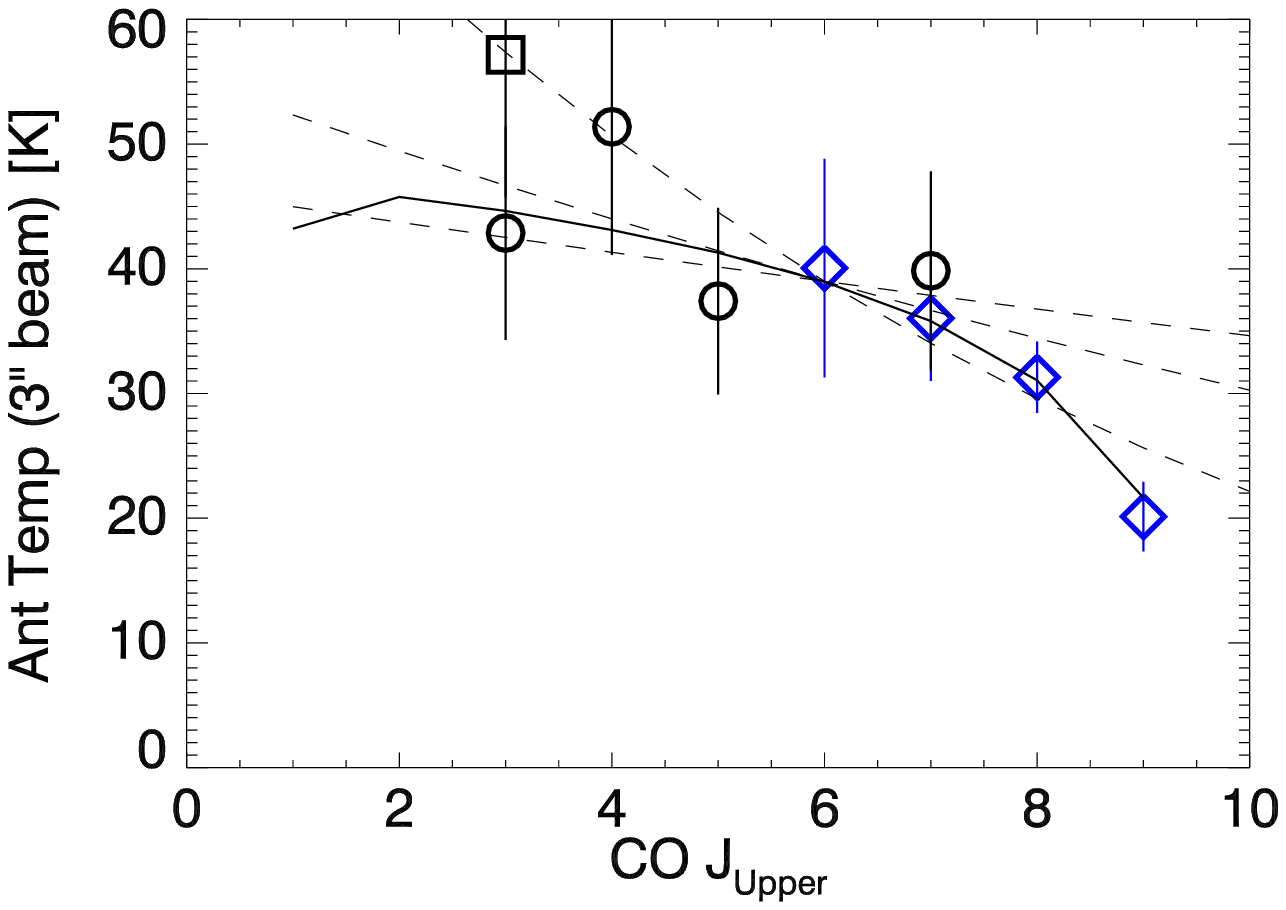}
\caption{Cloverleaf CO spectral line energy distribution (SLED) using all available transitions.  The top panel is in bolometric energy units, the bottom in brightness temperature units, referred to the VS03 source size assuming $m=11$.  We have adoped a 20\% systematic uncertainty for the measurements other than those from Z-Spec, shown in the temperature plot.    A model corresponding to conditions near the peak in the pressure likelihood  ($\rm T=56\,K, n_{H_2}=2.8\times 10^4, cm^{-3}$, $\rm N_{CO} / dv = 1.6\times10^6\,\rm cm^{-2}\,km^{-1}\,s$, normalized to match the observed \jeight\ flux) is overplotted as a solid line in both scales.
Dashed lines in the temperature scale show themalized blackbody emission for temperatures of 25, 50 and 100 K (higher T is shallower slope), arbitrarily normalized to the model value of 39~K at \jsix.  To produce 78\% of the total observed intensities, the area filling factors relative to the VS03 disk would have to be 3.3, 1.1, and 0.46, respectively.
\label{fig:sed}}
\end{figure}

\subsection{CO Excitation and Radiative Transfer Modeling} \label{sec:rad}
The run of CO line luminosity with $J$ for all published measurements is shown in Figure~\ref{fig:sed}.    
The luminosity scale shows the quoted statistical uncertainties for the various measurements.  Systematic errors are clearly present judging from the published \jthree\ through \jfive\ measurements, and we have assumed that systematic effects produce a normally-distributed uncertainty of $\sigma=20\%$ of the measured flux for the non-Z-Spec measurements.  This statistical uncertainty is shown in the temperature scale.

Our approach is to consider the properties of the gas which is confined to the physical size inferred by VS03, a 650~pc-radius disk inclined by $i=30^\circ$ to present a 559~pc semi-minor axis.   VS03 resolve 78\% of the total \jseven\ flux in this source, and we adopt this fraction for other mid-J transitions, in particular for \jeight, our adopted reference transitions for bolometric calculations given its clean measurement with Z-Spec.   The large observed line luminosities arising from the VS03 disk immediately imply that the molecular gas in the disk is very warm.   If the disk radiates isotropically with a velocity FWHM of 500~\kms,the brightness temperatures must be $\rm T_{RJ}=$  37, 40, 31, 20~K, for the \jsix, \jseven, \jeight, and \jnine\ lines, respectively.   The corresponding physical temperatures assuming optically-thick, thermalized emission are even higher: 50, 56, 49, 39~K.  An area filling factor or optical depth less than unity would correspond to even higher physical temperatures.   Adopting a velocity width smaller than the 500~\kms\ that we assume would also correspond to even higher physical temperatures.

For a more quantitative analysis, we begin with a variant of the RADEX code \citep{VanderTak_07} to
model the CO excitation and radiative transfer. RADEX is a non-LTE code that uses an iterative, escape probability formalism to treat the line radiative transfer. 
Three independent physical variables provide the input into RADEX: the gas density
($n_{\rm H_2}$), the kinetic temperature ($T$), and the CO column density
per unit linewidth $N_{\rm CO}/\Delta v$, which sets the optical depth
scale. RADEX calculates the excitation temperatures, line optical
depths, and line surface brightnesses in the CO lines.  Because of the uncertainties in the magnification and the beam filling factor (assumed to
be the same for all transitions), we compare the line ratios (with respect to the \jeight\ line), rather than
the absolute line fluxes. We have used RADEX in this way to generate the CO line intensities on a large grid in
density (10$^2$--10$^8$ cm$^{-3}$), temperature (0--300~K), and CO column per unit linewidth (10$^{14}$--10$^{20}\,\rm cm^{-2}\,km^{-1}\,s$).   

The CO column density per unit linewidth is not strongly constrained by the $^{12}$CO line ratios alone, though very small optical depths are not allowed because some degree of radiative trapping is required to populate the levels up to $J=9$ (T= 249~K).   To address this,  we also compare the upper limits for the $^{13}$CO \jseven\ and \jeight\ transitions in our band with the output of a RADEX model for this species, assuming that its fractional abundance relative to $^{12}$CO is 1/40, a value derived in a multi-level study of the NGC~253 \citet{Henkel_93}.  

We then follow the methodology of \citet{Ward_03} to generate
likelihood distributions for $n_{\rm H_2}$, $T$, and $N_{\rm
CO}/\Delta v$ by comparing the RADEX results with the observed
line ratios.   We assume uniform priors in the logarithm of all physical parameters, except for two prior constraints.   First, the total mass of gas producing the CO lines in the disk cannot exceed the dynamical mass implied by the observed velocity spread and the size of the disk.  This eliminates very large CO optical depths according to:  \begin{equation}
\frac{N_{\rm CO} }{dv} < \frac{M_{\rm dyn} X_{\rm CO}}{1.4 m_{\rm H_2}} \frac{1}{\pi R^2_{\rm d} \cos{i}} \frac{1}{\Delta v}= 1.3\times10^{18}\,\rm cm^{-2}\, km^{-1}\, s.\end{equation} 
Here the dynamical mass in the disk is based on the 650~pc modeled size, the measured aspect ratio implying $i=30^\circ$, and a circular velocity of 375\kms/$\sin{30^\circ}$, per the B97 HWZI profile width (and confirmed by W03):  $\rm M_{dyn}=8.5\times10^{10}\,M_{\odot}$.  $\rm X_{CO}$ is the CO abundance relative to \hh\ (taken at $2\times10^{-4}$) and $\mu=1.4$ is the mean molecular weight in units of $\rm M_{H_2}$.

\begin{figure}\centering
\includegraphics[width=8.7cm]{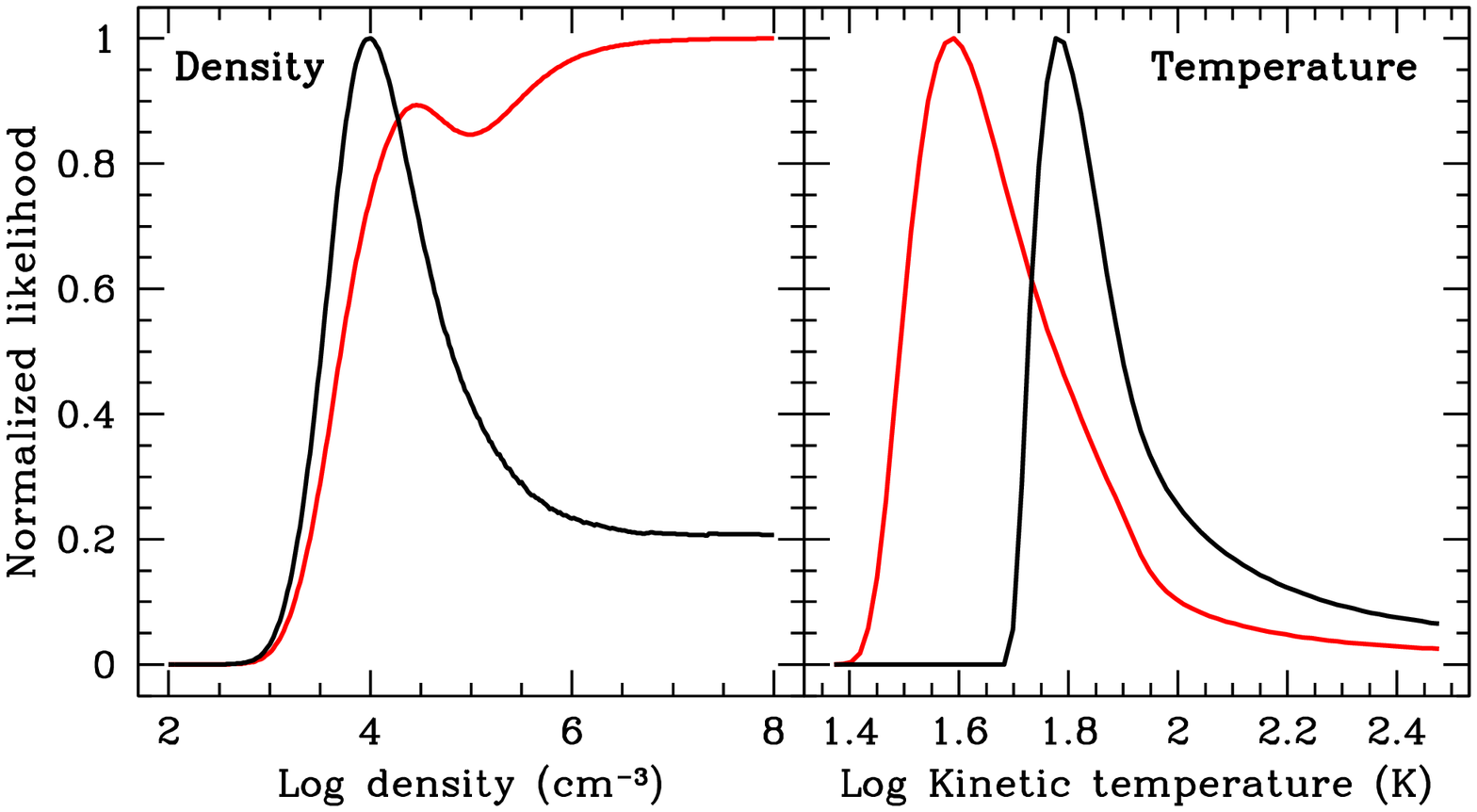}\\
\includegraphics[width=8.7cm]{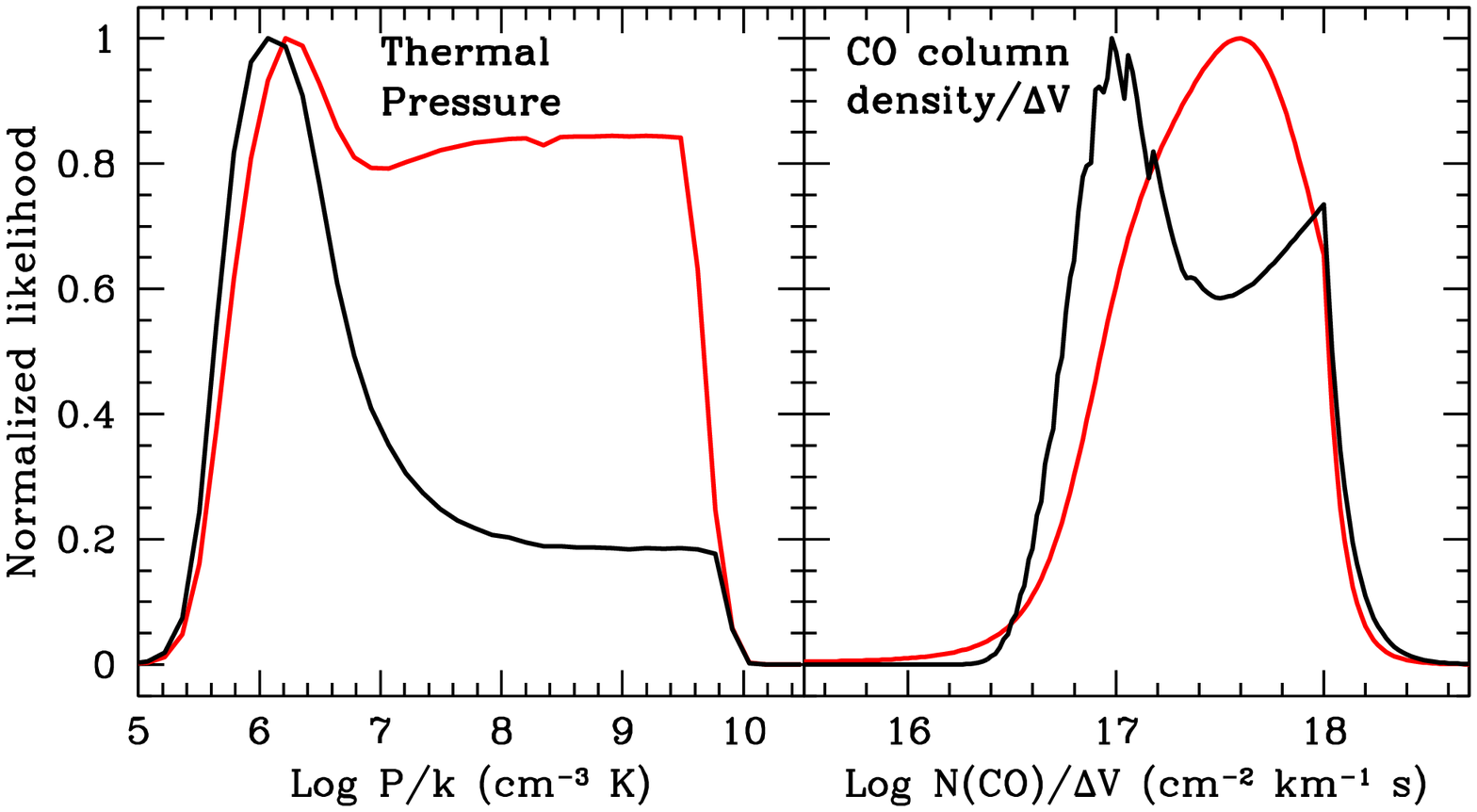}
\caption{Likelihood distributions for the physical conditions in the Cloverleaf host, with (red) and without (black) a prior applied to insure that the temperature is sufficient to produce the observed luminosity in region with size given by the modeled disk.  For both, the constraint in column density per linewidth is applied, eliminating N$_{CO} / \Delta v$ values much above $10^{18} \rm cm^{-2}\,km^{-1}s$.   Otherwise, uniform priors are adopted in the logarithm of n, T and N$_{CO} / \Delta v$.   All transitions are included.
   \label{fig:nt1}}
\end{figure}

The second constraint requires a minimum temperature sufficient to produce the observed luminosity in the finite-sized disk, described above.  Figure ~\ref{fig:nt1} shows the effect on the likelihood distributions of adding this constraint:  with it, the physical temperature is required to be above $\sim$50~K, without it, the temperature likelihood suggests T$\sim$30--60~K.   The constrained-temperature likelihood suggests somewhat lower densities, and provides a better-defined total thermal pressure than the unconstrained likelihood, peaking between 0.8--3$\times10^{6}\rm \,K\,cm^{-3}$.     We note that the derived likelihoods depend on the assumed linewidth only through this second constraint.  In its absence, the likelihoods depend only on the integrated intensities.   As described above, the adoption of 500~\kms\ linewidth value is actually a conservative value in this context: smaller linewidths corresponds to even higher temperatures, which would provide a more stringent constraint on the final likelihoods.
 
\begin{figure}\centering
\includegraphics*[width=8.7cm]{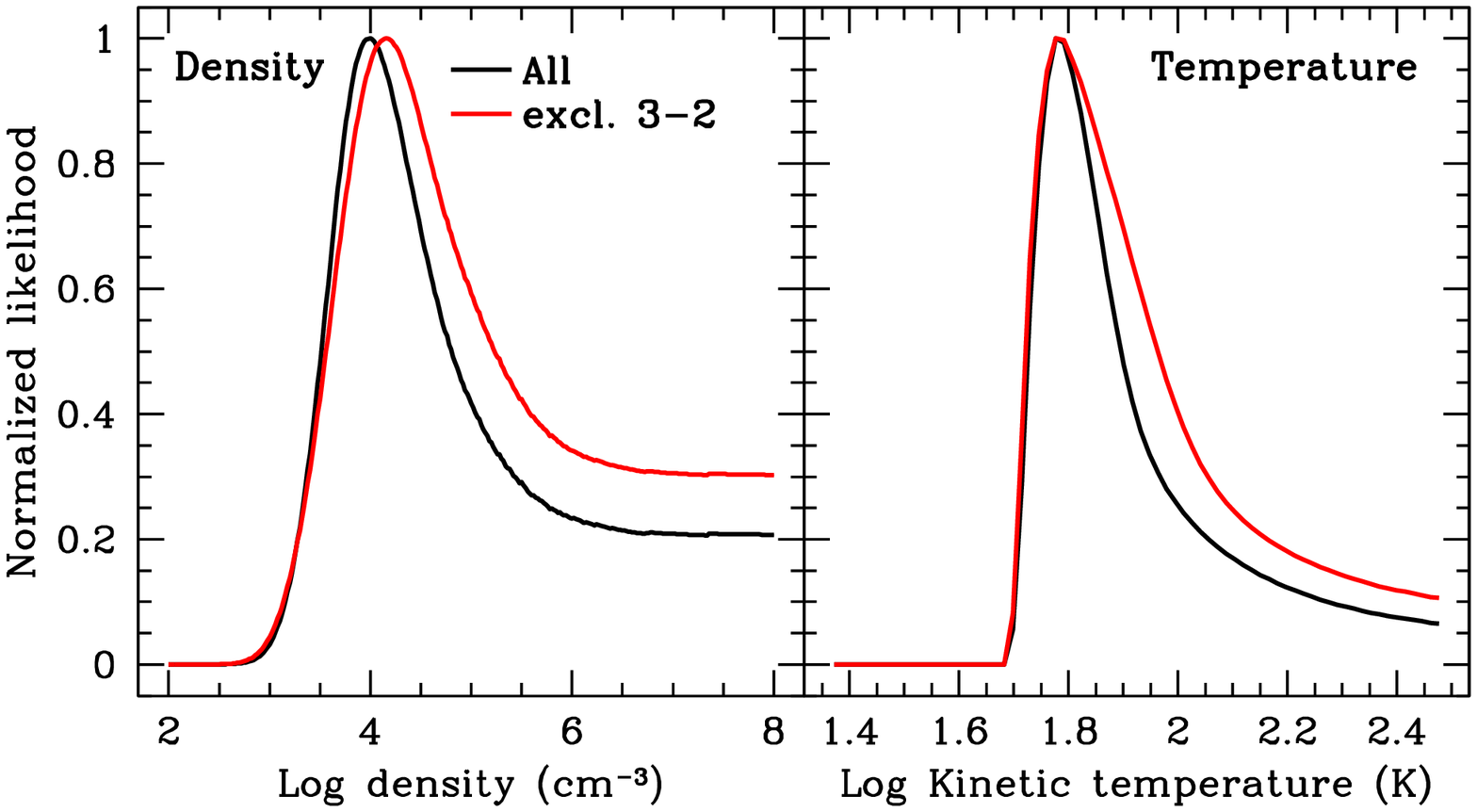}\\
\includegraphics*[width=8.7cm]{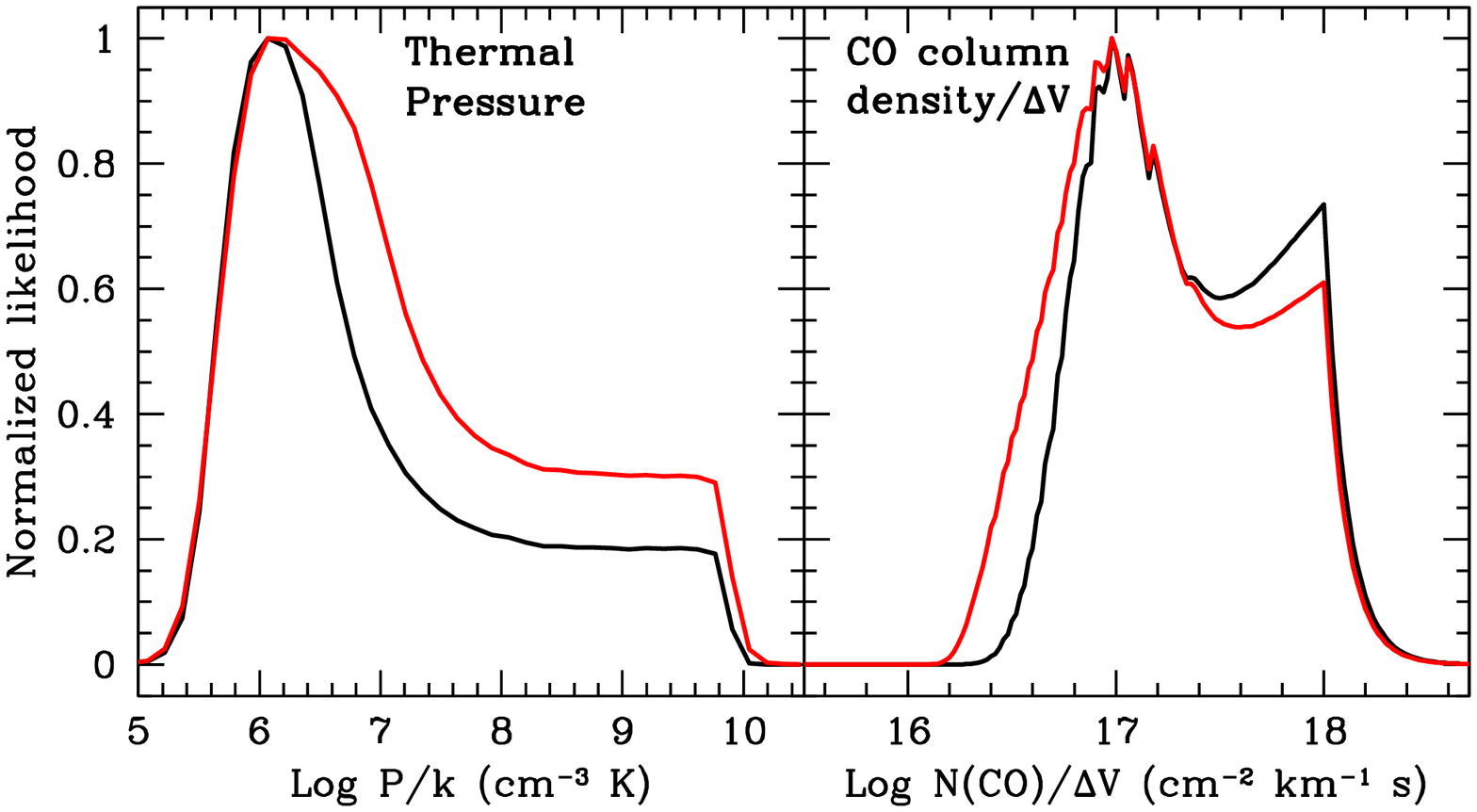}
\caption{Likelihood distributions for the physical conditions in the Cloverleaf host, both with (black) and without (red) the \jthree\ measurements included.  The temperature and optical depth constraints are applied, but otherwise uniform priors are assumed in the logarithms of n, T and N$_{CO} / \Delta v$. \label{fig:nt2}}
\end{figure}

While the disk dominates the mid-J line emission, it may not dominate the integrated large-beam low-J (e.g. \jone\ through \jthree) fluxes.  We explore the degree to which a single component model can account for all the transitions by both including and excluding the \jthree\ measurement (the lowest-J transition available).
 The results are shown in Figure~\ref{fig:nt2}.   Not surprisingly, higher excitation conditions are favored when the \jthree\ is excluded, but the effect is small, and indicates that a single component is a suitable model for all the observed transitions.  Our adopted likelihoods for the physical conditions are thus shown with the black curves in both Figures~\ref{fig:nt1} and \ref{fig:nt2}, yielding a temperature of 50--100~K and a density n$_{\rm H_2} > 3\times10^3\,\rm cm^{-3}$.  As an example, the spectrum in Figure~\ref{fig:sed} is overlaid with the fluxes predicted by a model near the peak of the pressure likelihood with $\rm T =  56\,\rm K$, $n_{\rm H_2}=2.8\times10^4\,\rm cm^{-3}$, and $N_{\rm CO}/\Delta v= 1.6\times10^{16}\,\rm cm^{-2}\,km^{-1}\,s$.   We note that for our peak likelihood conditions, the gas density and the optical depth parameter are consistent with the gas having at least enough velocity dispersion to correspond to virialized motion under its own self-gravity.  Rearranging Equation~2 in \citet{Papadopoulos_07} for our units gives: 
\begin{equation}
\rm K_{\rm vir} = \frac{19.0}{\sqrt{\alpha}}\left[ \frac{N_{\rm CO}/ dv}{10^{17}}\right]^{-1}\,\left[\frac{n_{H_2}}{10^3 cm^{-3}}\right]^{1/2}\,\left[\frac{X_{CO}}{4\times10^{-4}}\right] > 1 
\end{equation}
where $\rm N_{CO} / dv$ is in units of $\rm cm^{-2}\,km^{-1}\,s$, $\rm X_{CO}$ is the CO abundance relative to \hh, $\alpha$ is a parameter ranging from 1--2.5 per the cloud density profile, and the inequality accounts for the fact that the gas can of course be subject to more than its own gravity.   For a density of $10^4\,\rm cm^{-3}$, the inequality is satisfied as long as $\rm N_{CO} / dv$ is less than $3.8\times10^{18}\,\rm cm^{-2}\,km^{-1}\,s$.  Our derived likelihoods thus imply velocity dispersions which exceed the virial requirement by about an order of magnitude, perhaps due to large-scale turbulence and/or the influence of additional mass (of at most $\sim$3 times that in the gas itself).

\subsection{Mass of Molecular Gas}
Our results are broadly consistent
with the analysis of B97; cf. their Figure 2.  As in
that paper,  we can directly compute (a likelihood for) the mass by scaling from the observed line luminosities and using the implicit information on the line optical depth from RADEX, instead of relying on an uncertain CO X-factor
The calculation depends somewhat on the cloud geometry, as it determines the ratio of mass to emitting area; we have assumed spherical clouds, for which the working relationship is \begin{equation} \frac{M}{L}=\frac{c}{12\pi I}\frac{N_{\rm CO}/\Delta v}{X_{CO}}\,\mu\, m_{H_2} = 69\, N_{17}\left(\frac{X_{CO}}{10^{-4}}\right)^{-1}I^{-1} \,\frac{M_\odot}{L_\odot},  \end{equation} where $N_{17}=[N/\Delta v]/[10^{17}\rm cm^{-2}/(km\,s^{-1})]$, the intensity $I$ is in cgs integrated intensity units (erg s$^{-1}$ cm$^{-2}$ sr$^{-1}$).   This conversion results in a similar mass when applied to the various mid-J transitions, and Figure~\ref{fig:mass} shows the likelihood distribution for the total gas mass, again with and without \jthree. 
Our likelihood corresponds to a true molecular gas mass in the source of $M\sim 0.2-30 \times 10^{10} M_\odot$, 
%
consistent with previous gas mass estimates derived from CO, dust, and neutral carbon which range from 1.3--1.6$\times10^{10}$~\ms\ (W05).   We note for completeness the consistency around the peak values in the mass and optical depth likelihoods. The gas column density is the estimated $\rm M=6\times10^{9}\, M_\odot$ distributed over the (projected) VS03 disk: $\rm N_{H}\sim4.6\times10^{23}\,cm^{-2}$.  The CO optical depth parameter should be the corresponding CO column ($\rm N_{CO} = 2\times10^{-4}\, N_{H}\sim4.6\times10^{19}\,cm^{-2}$) divided by the linewidth of the source (400--500~\kms), giving $\rm N_{CO} / dv \sim 0.9-1.2\times10^{17}\,cm^{-2}\,km^{-1}\,s$.

\subsection{Origin of the Warm Molecular Gas}

\begin{figure}\centering
\includegraphics*[width=8.7cm]{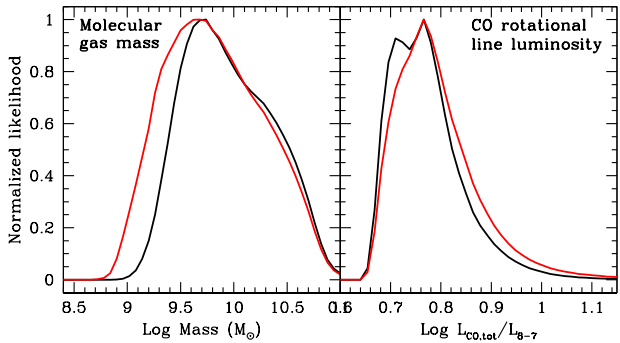}
\caption{LEFT:  Likelihood for the total molecular gas mass traced with CO in the Cloverleaf host, assuming a magnification factor $m$ of 11.   RIGHT:  Likelihood distribution for the total luminosity in the all the CO transitions.   Color coding is as in Figure~\ref{fig:nt2}}
\label{fig:mass}
\end{figure}

The CO emission observed in the Cloverleaf is unusually intense, even if the entire far-IR luminosity is produced by a starburst with $L_{\rm SB} = L_{\rm 40-120 \mu m} \approx 5.5\times10^{12}\,$\ls\, as suggested by \citet{Lutz_07}, W03.  The most luminous CO lines, the \jeight, \jnine\ each emit a fraction $\sim 10^{-4}$ of this total starburst luminosity (see Table~\ref{tab:lines}), and we measure the total CO luminosity through the likelihood formalism, finding that it is well-constrained at 4.6--6.8 $\times$ the \jeight\ luminosity (Figure~\ref{fig:mass}).   We therefore have
\begin{equation}
L_{\rm CO}\approx 3.3\times 10^{9}\rm\, L_{_\odot} \approx 6.1\times 10^{-4}\, L_{far-IR},
\end{equation}
assuming that the same value of $m$ applies to both the gas and dust.  
This is more extreme than for the nuclei of the nearby starburst galaxies.   In NGC~253, for example, the CO SED (in energy units) in the central 180~pc appears to peak at \jsix\ or \jseven; and each of these lines carry 2--3$\times10^{-5}$ times the far-IR luminosity, so the total CO luminosity fraction is $\sim$1--2$\times10^{-4}$ \citep{Bradford_03,Hailey-Dunsheath_08,Guesten_06}.   
Moreover, recent mid-J CO observations with the ZEUS spectrometer \citep{Stacey_07} in a sample of local-universe LIRGS and ULIRGS show mid-J (\jsix\ and/or \jseven) CO fractional luminosities comparable to or less than those in NGC~253 (T. Nikola, personal communication).  Thus molecular gas cooling relative to the far-IR dust emission in the Cloverleaf host demonstrably exceeds that in even the extreme local-universe systems.  We now consider potential heating sources for this bulk of warm molecular material.

\subsubsection{Stellar Ultraviolet (PDRs)}\label{sec:pdr}
 
Recent PDR model calculations (\citet{Kaufman_06}, hereafter K06) compute CO line intensities which can be compared directly with the observations.   These models parametrize the space of physical conditions with two quantities:  the incident far-UV flux ($\rm G_0$), and the molecular hydrogen density of the nascent molecular cloud ($n_{\rm H_2}$).   They indicate that for relatively high densities ($n > \rm few\times10^4$) and modest UV fields ($1<G_0<10^4$), the CO \jsix\ and \jseven\ transitions can have intensities more than $10^{-4}$ times the total far-IR luminosity.  We compare two measured ratios with model calculations available online as part of the K06 model:
1) the measured intensity ratio of the CO \jseven\ to the total far-IR ($=8.8\times10^{-5}$, assuming that they come from the same physical region), and 2) the CO \jsix\ to \jtwo\ (=23, assuming \jtwo\ is thermalized at the same temperature a s\jthree).   We find that these ratios can be consistent with the K06 PDR model with G$_0$=1--5$\times 10^3$ and $n_{\rm H_2} =$1--4$ \times 10^5\,\rm cm^{-3}$.   The density estimate, when combined with our likelihood analysis of the thermal pressure, suggests gas temperatures in the range of 10--300~K, reasonable given the range of temperatures expected in the PDR.

However, the neutral carbon intensities reported thus far are are not consistent with this picture.   For the PDR parameters discussed above, the K06 model predicts that the \ci \jone\ intensity relative to the CO \jthree\ and \jfour\ intensities should be 0.2--0.4 and  0.1--0.3, respectively; ratios which are close to those observed.   But the \ci\ \jtwo\ to \jone\ intensity ratio is predicted to be $\sim$6, while it is observed at only 2--2.5.   The PDR model is predicting warmer neutral carbon than is observed, a fact which is generally difficult to reconcile with the conditions derived from the CO spectrum and which is discussed further in \ref{sec:xdr} below.

For both total far-IR flux, and CO \jseven, the observed intensity emerging from the disk is $\sim$50 times greater than predicted by the single PDR model which fits the ratios.  If UV photons are responsible for the heating of the gas, then on average in the Cloverleaf disk, we are seeing $\sim$50 PDR surfaces superposed along the line of sight.   Correspondingly, the column density corresponding to a visual extinction ($\rm A_{\rm V}$) of 10 magnitudes is $\sim$1/50 of the total gas column estimated in Section~\ref{sec:rad}, if it is averaged over the 650-pc disk.   The derived far-UV field of $G_0 \sim $1--5$\times 10^3$ is modest, suggesting that the Cloverleaf PDR is not directly adjacent to an OB star as in an ionization-bounded HII region, but that the PDRs are typically 0.1 to 0.5 pc from the OB stars.   This is quite reasonable, if the $5.5\times10^{12}$ \ls\ attributed to the starburst arises from 100--10000 \ls\ stars in a region of size given by the VS03 model, then the average interstellar separation is 0.2--1~pc.

\subsubsection{Hard X-Rays from the Active Nucleus}\label{sec:xdr}
The hard X-ray continuum of the active nucleus itself could power the observed CO
emission.  Hard (E$>$1~keV) X-rays can penetrate a large gas column (N$_{\rm H}>$10$^{22}$~cm$^{-2}$) forming an X-ray Dissociation Region (XDR:  \citet{Maloney_96}).  The key quantity for understanding the XDR heating and cooling is the ratio of the X-ray energy deposition rate per particle (set by the X-ray luminosity and attenuating column) to the gas density which sets the recombination and cooling rates.  We do not have a direct view of the AGN in the Cloverleaf.  The observed X-ray emission is believed to be scattered \citep{Chartas_00}, and even this scattered emission is subject to an uncertain absorption in the gas along the line of sight.  Thus the true X-ray flux must be inferred or modeled based on the observed spectrum.   At the bolometric luminosity estimated for the Cloverleaf AGN ($\rm L_{\rm bol}\approx 7\times
10^{13}$\ls \citep{Lutz_07}), the $1-20$
keV X-rays typically carry $\sim 5\%$  of $L_{\rm bol}$ \citep{Mushotzky_93} (though with a scatter of a factor of $\sim 3$ in this
relation). This implies that $\rm L_x\approx 3\times 10^{12}$\ls.  
\citet{Chartas_07} present a model (their Figure~9) explaining the Cloverleaf X-ray spectra from Chandra and XMM-Newton in which the total 2--10 keV luminosity of the central region is $2\times10^{46}\,\rm erg\,s^{-1}$ or 5$\times 10^{12}$\ls (instrinsic, with $m\sim10$), based on an observed flux which is a factor of $\sim$200 lower.

While these inferences clearly have large uncertainties, it is very likely that the hard X-ray luminosity approaches the 40--120\mm\ far-infrared luminosity, and
$$
{L_{\rm CO}\over L_{\rm X}}\approx 10^{-3}.
$$
The large column densities derived in the CO-line analysis mean that much of this hard X-ray emission will be absorbed.   Unlike in a PDR, the X-rays input a much greater fraction their energy ($\sim$0.1--1) into the gas than the dust, and since the CO lines will be an important coolant, this ratio immediately indicates that the AGN could readily power the CO emission we observe.

To explore this possibility, we have generated XDR models for
parameters appropriate to the Cloverleaf system. The input parameters are the
total gas density and the incident X-ray flux, which is set (for fixed
$L_{\rm X}$) by the distance to the AGN and the column density of
absorbing gas between the hard X-ray source and the
modeled region. The results also depend on the optical depth in the coolings lines; we have used a total hydrogen column per unit linewidth
of $10^{21}\rm cm^{-2}/km\,s^{-1}$, as suggested by the CO line
analysis.

\begin{figure}\centering
\includegraphics*[width=8.5cm,bb=23 10 490 410]{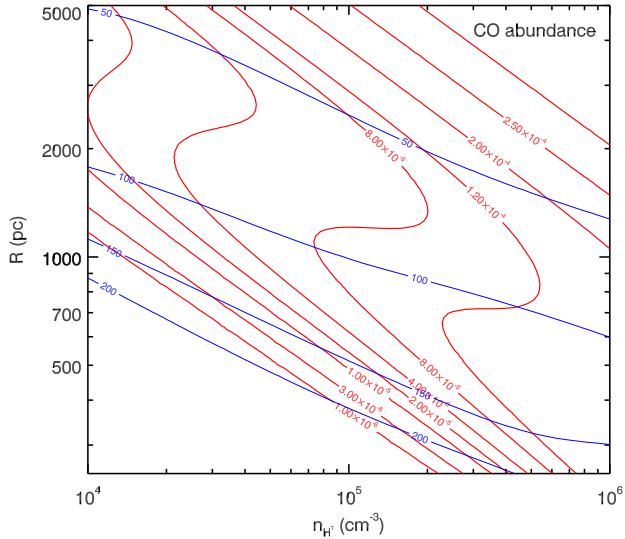}
\caption{CO abundance (relative to H nuclei) in the distance-gas density space of the XDR model.  At small distances and low densities, gas is largely atomic.  The blue contours show gas temperature. \label{fig:X1}}
\end{figure}

Figure~\ref{fig:X1} shows the predicted CO abundance as a function of gas density and cloud distance (denoted R) from the nucleus.  The assumed X-ray luminosity is conservatively adopted to be 10$^{46}\,\rm erg\,s^{-1}$ and the total attenuating
column is $\rm N_{H, att}=3\times 10^{23} cm^{-2}$, of order but less than the estimated total column from the CO analysis.  This results in a factor of 22 attenuation and the resulting X-ray flux at 600~pc distance is thus 10~$\rm erg\,s^{-1}\,cm^{-2}$.  Larger values of L$_{\rm X}$ or smaller values of $N_{att}$ shift the
contours up and to the right.  We have adopted solar abundances for elements in the gas-phase.   At small R and relatively low
density (lower left corner of the plot), the CO abundance is small, as
the gas is warm and atomic.  However, over most of the plotted range,
the CO abundance is large (more than $\sim$a few $\times 10^{-5}$),
reaching nearly $3\times 10^{-4}$ at the largest R and
densities, for which all the gas-phase carbon is in CO.  

\begin{figure}\centering
\includegraphics*[width=8.5cm]{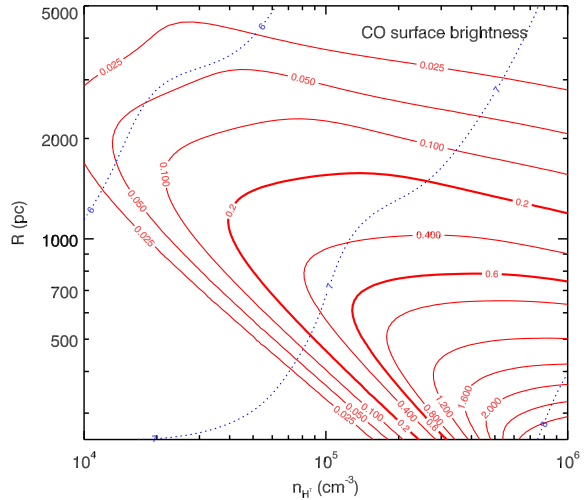}
\caption{
CO rotational cooling for X-ray heated clouds in the XDR model space.   The contours show surface brightness of all the CO lines in units of erg / s / cm$^2$.  Values of 0.2 and 0.6 correspond to the observed surface brightness assuming filling factors between 1/3 and 1 in the VS03 disk.  Dashed blue contours show thermal pressure, values denoted with $\log{\rm nT}$ \label{fig:X2}} 
\end{figure}

As noted above, the surface brightness of the CO emission in the
Cloverleaf is very large. The estimated total CO
rotational line luminosity (equation [2]), distributed over the VS03 disk with unit area filling factor
implies a surface flux of approximately 0.2 ergs cm$^{-2}$ s$^{-1}$.   If the area filing factor is less than unity, then even larger local surface brightness is required. 
We plot this local CO surface brightness predicted by our XDR models
in Figure~\ref{fig:X2}, and conclude that for densities above
$3\times10^4$ cm$^{-3}$, and R from 500--1500~pc, our model is consistent with the observed CO cooling, and suggests area filling factors in the range of 0.3--1.   Of course, even if the area filling factor is unity, we note that the {\it volume} filling factors can still be quite small.   Our results are consistent with the XDR model of \citet{Meijerink_07}, who present the CO \jseven\ to \jthree\ intensity ratio as a function on incident X-ray flux and density (their Figure 6).   The observed ratio is 8.9--12.3, depending which\jthree\ measurement is adopted, suggesting an X-ray flux of 5--20~$\rm erg\,s^{-1}\,cm^{-2}$, which compares well with our working value of $\sim$10 at $\rm d=600\,pc$.

\begin{figure}\centering
\includegraphics*[width=8.5cm,bb=23 10 490 410]{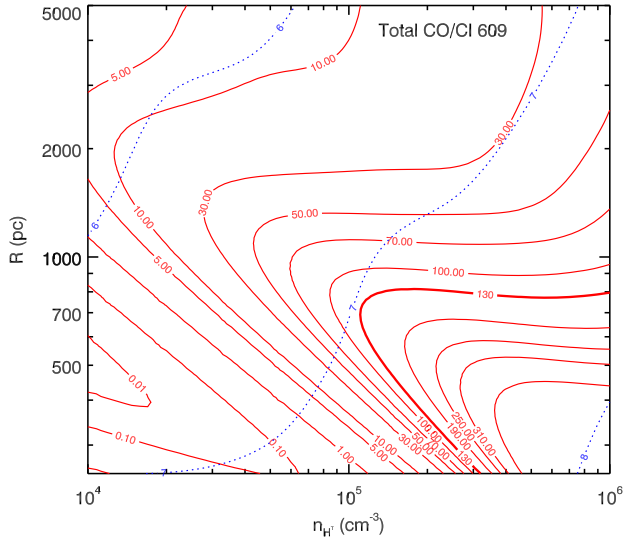}
\caption{Ratio of the total CO cooling to the \ci\ \jone\ cooling in the XDR model.   The heavy contour corresponds to the observed value (130).  \label{fig:ci}}
\end{figure}

More generally,
the XDR models indicate that CO rotational line emission can be $\sim
0.1$ of the absorbed X-ray energy, so only $\sim 1\%$ of the hard
X-ray luminosity of the AGN needs to be absorbed in a CO-luminous XDR
to produce the CO emission.   However, we emphasize that while the XDR can easily heat the gas, it is not likely to appreciably heat the co-extant dust the way that the PDR front would.     
In an XDR, the ratio of total CO rotational line flux to the {\it locally-generated} far-infrared flux
from grains (due to absorption of emitted line photons and to UV
photons produced by excitation and re-combination) can easily exceed
$10^{-3}$, and can be as large as $\sim 0.1$.   (This ratio depends on
the cloud column density as well as the column density per unit
linewidth, since the far-infrared dust emission depends on the former.)
Since the observed ratio is $\sim 5\times 10^{-4}$, the observed far-IR / submillimeter dust emission can not be powered by the XDR processes.  There must be additional source(s)
of local grain heating, e.g., star formation, or re-radiated continuum
from the AGN.   

As with the PDR, the XDR framework does not perfectly account for the observed \ci\ intensities, though it may provide a better match in terms of total cooling.  The XDR model generally predicts lower \ci\ relative to CO than the PDR, and the \ci\ \jone\ line is well-matched with the XDR model.  The ratio of the total CO emergent intensity including all transitions to the \ci\ \jone\ intensity is plotted in Figure~\ref{fig:ci}.  The observed value of 130 is highlighted, as are ratios a factor of two above and below this value to allow for systematic uncertainties in both.   As with the analysis described above, the allowed region corresponds to sub-kpc distances and densities greater than 10$^5\,\rm cm^{-3}$.  However, it remains true that the \ci\ temperature is far too low; the \ci\ \jtwo\ to \jone\ ratio (R = 2.3 in power units per W03, W05) suggests a physical temperature of only $\sim$30~K, a value generally inconsistent with both the  the XDR and the PDR models.   W05 suggest that the excess \ci\ \jone\ emission may arise in an (additional) cool gas component exterior to the material traced with the CO \jseven.  Such a scenario may be consistent with our findings, as we allow for a second component of low-excitation material outside the disk.  However, this would mean that the \ci\ \jone\ which is associated with the mid-J CO would be less, corresponding to higher ratios in Figure~\ref{fig:ci}, requiring higher densities and lower distances from the central source.  A large allocation of the \ci\ \jone\ to this external component is difficult to reconcile with the overall XDR picture.  Another possibility is that the measurements of the relatively weak \ci\ intensities are subject to systematic errors.  We do note that our measurement favors a slightly higher flux for the \ci\ \jtwo\ than W05 (giving $R\sim3.8$, for T$\sim$40~K), though with low significance.  We do note that the W05 measurement of \ci\ jtwo puts the \ci\ ratio well below that of M82 (R = 4.3, per \citep{Stutzki_97}), the only external galaxy for which the \jtwo\ measurement is published. 

In the context of the XDR model, the above analyses suggest that the gas density is high: $\rm n_{H_2} \ge 5 \times 10^4 cm^{-3} $, for the bulk of the material.  This density is higher than required by the CO intensity ratios alone (Figure~\ref{fig:nt2}); it derives in essence from requiring the total observed CO luminosity to be generated in a volume constrained by the VS03 disk model.    High densities have been suggested given the measurement of high-dipole-moment species, beginning with HCN \jone\ (\citet{Solomon_03}), and now with HCO$^+$ \jone\ \citep{Riechers_06}.   The two transitions are comparable: $\rm L_{HCO^+} = 0.8 \pm 0.2 L_{HCN}$, and the fractional intensity relative to L$_{\rm FIR}$ nicely matches the linear correlation found by \citet{Gao_04} for the local-universe LIRGS and ULIRGS.   While mid-IR pumping schemes could potentially account for these lines, \citet{Riechers_06} conclude that the match between the HCN and HCO$^+$ lines is best explained by both species producing optically-thick thermalized emission (at least for the \jone\ transition), requiring $\rm n_{H_2} \ge 10^5$.  This is fully consistent with the both the XDR and the high-density PDR scenarios.

\subsubsection{Turbulence, Cosmic Rays}\label{sec:turb}

Finally, we note that there are other potential energy sources which will preferentially heat the gas relative to the dust.  Cosmic-ray-ionization heating has been proposed to explain the warm molecular material in the nuclei of M~82 and NGC~253 \citep{Hailey-Dunsheath_08, Bradford_03, Suchkov_93}.   Cosmic Rays are generated in supernovae associated with massive star formation, and like X-rays, they penetrate throughout the bulk of the molecular material.  Simple dissipation of mechanical energy is also a potential heating source.  \citet{Bradford_05} apply the shock models of \citet{Draine_83, Draine_84, Roberge_90} to the molecular gas around SGR~A$^*$ and find that low-velocity ($v\sim$10--20~\kms) magnetohydrodynamic (MHD) shocks produced by cloud-cloud collisions can produce warm material which cools primarily via the mid-J CO rotational transition, again without appreciable heating of the dust.  The arguments presented in these studies would apply to the Cloverleaf host, and could also boost the mid-$J$ CO emission relative to that which arises in the PDR alone.     

\subsubsection{Prospects for Atomic Gas Tracers}

Further distinction between the PDR scenario and the X-ray and other various bulk heating scenarios could be provided by the atomic fine-structure transitions.  
In particular, the [OI]~$63\mu$m and [CII]~$158\mu$m will be important.  The ratio of these fine-structure line intensities to the total CO cooling is a powerful discriminant:  in the PDR the fine structure lines carry an order of magnitude more power than the CO since most of the UV heating is deposited in the atomic gas, while in the XDR, the energy is more uniformly distributed between the atomic and molecular components.   Indeed, our models suggest that the [OI]~$63\mu$m intensity and the total CO rotational cooling are comparable for the XDR.  With the Cloverleaf's unusual luminosity and lensing factor, these fine-structure lines may be accessible with the spectrometers on Herschel.

\subsubsection{Conclusion -- A Composite Solution}\label{sec:composite}
We conclude that there are several potential gas heating sources, each of which could contribute substantially to powering the CO emission in the Cloverleaf host.  The far-IR / submillimeter continuum component appears distinct from the mid-IR component, and the PAH fractional luminosities relative to this far-IR dust component are similar to those observed in starburst systems.  These facts are difficult to explain without a massive starburst.   On the other hand, since the fractional energy carried in the molecular gas exceeds that of the local starburst analogs by factors of 2--5, and there is a powerful X-ray source which is demonstrably capable of contributing 10--100\% of the required heating, it is difficult to exclude the X-rays as a heating source.   We conclude that the molecular material around the Cloverleaf QSO is most likely heated by both UV photons from young stars and and X-rays from the accretion zone, with neither dominant by more than $\sim$1 order of magnitude.  

X-rays which penetrate and heat the bulk of the molecular gas will impact the properties of the star formation, particularly the stellar mass function.   Simple thermodynamic arguments along the lines of the Jeans mass or Bonnor-Ebert mass show that the characteristic mass required for gravitational collapse increases as the 'minimum temperature' of the gas increases.  This is precisely the effect that a bulk heating mechanism would produce.   Neglecting the effects of magnetic fields and rotation, this mass is given by \begin{equation}
M_{BE} = \frac{C_{BE} \,v_T^4}{P^{1/2}\, G^{3/2}} \propto \frac{T^{3/2}}{n^{1/2}} \propto \frac{T^2}{P^{1/2}}, \label{eq:bonner}\end{equation} where $T$, $n$, and $P$ are temperature, number density, and thermal pressure, respectively of the environment from which the stars must form, and $C_{BE}$ is a numerical constant.  Detailed theoretical approaches confirm the sense of this relation, finding a scaling between the characteristic mass scale $M_*$ in a stellar IMF, and the minimum temperature to which molecular gas can cool---$M_{*} \propto T_{min}^{\gamma}$---with exponent $\gamma$ ranging from 1.7 (obtained in numerical experiments \citep{Jappsen_05}) to 3.35 (obtained in an analytic treatment \citep{Larson_85}).    In the Galaxy, $M_*\sim0.5$~\ms\ and $T_{min}\sim8$~K, values numerically consistent with Equation~\ref{eq:bonner}.   While our analysis does not directly measure $T_{min}$, it must be true that $T_{min}$ is increased relative to the Galaxy.   A plausible assertion is that $T_{min}$ scales as the typical temperature derived from fits to CO line ratios: at least $\sim$50K per our analysis of the Cloverleaf fluxes, compared with $\sim$22~K in the inner Galaxy per the COBE FIRAS measurements \citep{Fixsen_99}.    This simple scaling would suggest that $M_*$ in the Cloverleaf starburst is 4--15 times the Galactic value: some 2--5~\ms.   As \citet{Larson_98} and others have pointed out, such a top-heavy IMF converts a given mass of gas into a greater total stellar (and thus far-IR) luminosity than with the Salpeter IMF.   This is the leading scenario proposed to explain the 
factor of $\sim$3--5 discrepancy between the observed stellar mass buildup and the star-formation history in the $4<z<1$ era \citep{Perez-Gonzalez_08,Dave_08,Hopkins_06}.  While the Cloverleaf may be somewhat more extreme than the typical star-forming system, the increased prevalence of AGN in this early epoch may generally result in more massive stars than those formed in the Galaxy today.

\subsection{Additional Features in the Spectrum}

While it has yet to be explored completely in a extragalactic source, the short submillimeter band is expected to host low-lying transitions of light molecules and ions other than CO, some of which may be bright.  We now discuss some tentative identifications which could be followed up with the large collecting area and spectral resolution of the interferometers.

\subsubsection{Water in the Cloverleaf}\label{sec:water}

The feature at 277.6~GHz is fit with a Gaussian line of $\nu_{\rm rest} = 987.4 \pm 0.55\rm GHz$, adopting the systemic redshift of the Cloverleaf.   This might be identified with the lowest-lying transition of ortho-hydronium $o-\rm H_3O^+$ ($0^-_0\rightarrow 1^+_0$, $\nu_{\rm rest} = 984\rm\,GHz$), but unlike the other low-lying $\rm H_3O^+$ transitions ($\nu_{\rm rest} = 396, 388, 364\, \rm GHz$), it is not expected to be bright given the lack of radiative pumping pathways to excite the upper level \citep{Phillips_92}.  Moreover, the abundance of $\rm H_3O^+$ even in extreme Arp~220-like nuclei is believed to be less than $1/10$ that of water \citep{VanderTak_08}.    A better match to the fit, and a more convincing scenario is that this feature is the 987.9~GHz (\waterline) transition connecting the first excited level of $p$-\water\ at 54~K to the second excited level at 101~K.  Attempts to detect high-redshift rotational water transitions have been made.  \cite{Encrenaz_93,Casoli_94} report tentative detections of the \waterlinenon\ transition ($\nu_{\rm rest}=752.0\,\rm GHz$) in IRAS F10214+4724 ($\rm z=2.3$).  Remarkably, the reported intensity (0.65~\kkms\ at 30-m) is only a factor of 0.45 times that of the nearby CO \jsix\ transition (1.4~\kkms~30-m, per \citet{Solomon_92b}).  More recently, \citep{Riechers_06a} report an upper limit on the $3_{1,3}\rightarrow2_{2,0}$ (183~GHz) transition in MG~0751+2716 at $\rm z=3.2$.   

The a priori interpretation of water spectra is difficult even with multiple line detections, as evidenced by the work analyzing the ISO spectra toward SGR~B2 \citep{Neufeld_95,Cernicharo_06} and the Orion outflow \citep{Harwit_98, Cernicharo_06a}.  
Morever, since most of the low-lying water lines are not accessible at zero redshift from the ground, there are no good Galactic templates with which to compare the \waterline\ measurement.  Surveys have been conducted only in the ground-state \groundortho\ (538~\mm) line with the KAO, SWAS, and ODIN (e.g. \cite{Ashby_00, Neufeld_03, Snell_00}), finding low abundances ($<10^{-8}$) or upper limits in GMCs and cloud cores.   Similarly, \citet{Wilson_07} derive upper limits to the water abundance of $<10^{-8}$ from non-detections in nearby starburst galaxies, albeit with the large (2.1\arcmin) ODIN beam.  However, much larger water abundances (more than $10^{-5}$) are inferred in warm cores and outflows, notably the Orion outflow.

In spite of the ODIN ground-state non-detections in local starbursts, the ISO far-IR spectrum of the Arp~220 nucleus \citep{Gonzalez-Alfonso_04} demonstrates that water can be abundant and produce powerful features even in a large-beam (i.e. average) extragalactic spectrum, at least in extreme sources.  The 179~\mm\ absorption from the nucleus shows an equivalent width of $\sim$400\kms (in absorption).   These investigators quote \water\ column densities of 2--10$\times10^{17}\,\rm cm^{-2}$ toward the Arp~220 nucleus, corresponding to abundances of 1--5$\times10^{-8}$ for N$_{\rm H2}\sim 10^{25}\,\rm cm^{-2}$.   We measure an equivalent width of $\sim$580\kms, which is of the same order, albeit in emission.   We do note that the models of \citet{Cernicharo_06a} suggest that even in the SGR B2 geometry, the \waterline\ transition in particular could be brought into emission (in contrast with most of the other transitions) due to a favorable pumping / cascade network in the para levels.  While there is no straightforward detailed reconciliation of the tentative detection with published models, we note that water emission is more easy to understand in the context of an XDR than in a traditional PDR.   Our XDR model suggests a water abundance of order $10^{-7}$, greater than that derived in Arp~220, and more than would be expected in the bulk of molecular material of a PDR. 

\subsubsection{Upper limits, absorption identification}\label{sec:limits}
Before concluding, we briefly note for completeness upper limits for two other transitions, marked in the spectrum in Figure~\ref{fig:spec}.   The radical CH$^+$ has its ground state transition (\jone) at $\nu_{\rm rest} = 806\rm GHz$ ($\nu_{\rm obs}= 234.7\rm GHz$). The spectrum shows this channel is above the continuum at the 1.5$\sigma$ level; a formal 2.5$\sigma$ upper limit to the flux is 14~Jy~km/s.   The LiH \jtwo\ transition at $\nu_{\rm rest} = 887\rm GHz$ ($\nu_{\rm obs}= 249.4\rm GHz$) is also above the continuum at low significance, with a formal upper limit of 12.3 Jy~\kms.  

The absorption feature at 287~GHz is significant, at 3$\sigma$,  Interestingly, this frequency corresponds to the 557~GHz ground state of o-\water\ at z=0.93, perfectly within the redshift range expected for the Cloverleaf lens \citep{Kneib_98b} ($\bar{z} = 0.9 \pm 0.1$).  
The equivalent width of the absorption is $\sim$600 km/s, which would imply that there must be nearly complete absorption over a several hundred km/s velocity range.   A galaxy perfectly positioned along the line of sight could potentially do this.   For $\tau$ to be unity over 600 km/s requires a molecular hydrogen column of $6\times 10^{23}\,\rm cm^{-2}$, assuming an ortho-to-para ratio of unity, a water abundance of 10$^{-8}$, and that there is negligible population in the upper levels.  The lens is located 0.6 arcsec from the projected Cloverleaf images, thus the light passes within $<$5~kpc of the lens, so the geometry is not implausible.   Further investigation of this intriguing possibility will require higher spectral and spatial resolution measurements.

\section{Conclusions}

We have built and are now using the first broadband spectrograph for the millimeter / submillimeter, providing a new observational approach for studying galaxies at all redshifts.   Among the first experiments is the measurements of the rest-frame 272--444~\mm\ spectrum of the Cloverleaf system which contains the bolometric peak in the CO rotational spectrum.   In our analysis of the CO intensities and other published data we find the following:
\begin{enumerate}
\item{The intensity ratios of the \jthree\ through \jnine\ CO transitions can be produced with a single gas component.  The excitation is high, with thermal pressure $> 10^6\,\rm K\,cm^{-3}$.  The 0.2--5$\times 10^{10}$ \ms\ of excited molecular gas in the Cloverleaf host must fill a projected size of at least half the 650-pc-radius \citet{Venturini_03} disk.  A more compact distribution is not physical based on the absolute line intensities.}
\item{As \citet{Lutz_07} conclude, the match between the far-IR dust component and the PAH emission suggests that far-IR-emitting dust component represents a massive starburst.   However, the molecular gas cooling in the CO transitions alone is a large fraction ($6\times10^{-4}$) of the bolometric luminosity of this starburst.  This is a factor of a few larger than in the local-universe starburst galaxies for which mid-J CO intensities are available.}
\item{Given the powerful AGN of the Cloverleaf system, X-rays are likely to be an important energy source for molecular gas even several hundred parsecs from the nucleus.  Our XDR model show that $\sim$5\% of the bolometric luminosity of the AGN in hard X-rays can easily provide the energy input required to match the observed CO cooling.  The X-rays do not appreciably contribute to the far-IR dust continuum.}
\end{enumerate}
Our interpretation is that the Cloverleaf host is indeed a undergoing a massive starburst, but that it has additional energy input into the ISM via hard X-rays originating in the accretion zone.   As a result, the stellar IMF is likely biased toward much higher masses than in the Galaxy, with a typical stellar mass $M_*$ of several \ms.  With this characteristic mass replacing the Salpeter value of $\sim$0.5 \ms, the gas consumption rate at the observed luminosity is several times lower than that given by applying the \citet{Kennicutt_98} prescription.  The starburst we are witnessing in the Cloverleaf may therefore extend much longer than the 30~Myr estimated by \citet{Lutz_07}; it could last for a few hundred Myr.   A similar scenario, though perhaps less extreme, may be typical of the star-forming galaxies in the first half of the Universe's history when energy release peaked.

\acknowledgements{We are indebted to the staff of the Caltech Submillimeter Observatory for their help in Z-Spec's commissioning and observing.  We acknowledge Peter Ade and his group for some of the filters and Lionel Duband for the $^3$He / $^4$He refrigerator in Z-Spec, and are grateful for their help in the early integration of the instrument.   We benefitted from conversations with Tom Phillips, Paul Goldsmith, Simon Radford, and Andy Harris, as well as helpful comments from Xinyu Dai and an anonymous referee.  Finally, we acknowledge the following grants and fellowships:  NASA SARA grants NAGS-11911 and NAGS-12788, an NSF Career grant (AST-0239270) and a Research Corporation Award (RI0928) to J. Glenn, a Caltech Millikan and JPL Director's fellowships to C.M.B., a NRAO Jansky fellowship to J. Aguirre, and NASA GSRP fellowship to L. Earle. The research described in this paper carried out at the Jet Propulsion Laboratory, California Institute of Technology, was done so under a contract with the National Aeronautics and Space Administration. }

\bibliography{bradford_master}

\end{document}